\documentclass[epj]{svjour}
\newcommand{\sqrts}{\sqrt{s}}
\newcommand{\produc}{{\rm prod}}
\newcommand{\jpsi}{J/\psi}
\newcommand{\sqrtspsin}{\sqrt{s_{_{\jpsi {\rm N}}}}}
\newcommand{\mpsi}{m_{_{\jpsi}}}

\newcommand{\A}{{\rm A}}
\newcommand{\B}{{\rm B}}
\newcommand{\TA}{T_{_\A}}
\newcommand{\dd}{{\rm d}\,}
\newcommand{\ndf}{{\rm ndf}}
\newcommand{\sig}{\sigma_{_{J/\psi {\rm N}}}}

\newcommand{\sigb}{\bar{\sigma}_{_{J/\psi {\rm N}}}}

\newcommand{\sigsh}{\sigma^{\rm nPDF}_{_{J/\psi {\rm N}}}}

\newcommand{\sigEKS}{\sigma^{{\small {\rm EKS}}}_{_{J/\psi {\rm N}}}}
\newcommand{\sigEPS}{\sigma^{{\small {\rm EPS}}}_{_{J/\psi {\rm N}}}}

\def\X{{\rm X}}

\usepackage{graphics}
\usepackage{amsmath}
\usepackage{amssymb}

\begin{document}
\title{Global analysis of $\jpsi$ suppression in cold nuclear matter}
\author{Vi-Nham Tram\inst{1} \and Fran\c{c}ois Arleo\inst{2}
}
\institute{Lawrence Berkeley National Laboratory (LBL), 1 cyclotron Road, Berkeley, CA  94720-8169, USA \and Laboratoire d'Annecy-le-Vieux de Physique Th\'eorique (LAPTH), BP110, 74941 Annecy-le-Vieux, cedex, France}
\date{Received: date / Revised version: date}
%
\abstract{
Interpreting the $\jpsi$ suppression reported in nucleus--nucleus collisions at SPS and RHIC requires the quantitative understanding of cold nuclear matter effects, such as the inelastic rescattering of $\jpsi$ states in nuclei or the nuclear modification of parton densities. With respect to our former Glauber analysis, we include in the present work the new PHENIX $d$--Au measurements, and analyze as well all existing data using the EPS08 nuclear parton densities recently released. The largest suppression reported in the new PHENIX analysis leads in turn to an increase of $\sig$ from $3.5\pm0.3$~mb to $5.4\pm2.5$~mb using proton PDF. The stronger $x$-dependence of the $G^{\A}/G^p$ ratio in EPS08 as compared to e.g. EKS98 shifts the cross section towards larger values at fixed target energies ($x_2\sim0.1$) while decreasing somehow the value extracted at RHIC ($x_2\sim10^{-2}$).%
\PACS{
      {PACS-key}{discribing text of that key}   \and
      {PACS-key}{discribing text of that key}
     } 
} 
\maketitle
\section{Introduction}
\label{intro}
The suppression of heavy-quark bound states in heavy-ion collisions due to Debye screening  is known to be a sensitive probe for quark-gluon plasma (QGP) formation~\cite{Matsui:1986dk}. However, reactions involving heavy nuclei introduce ``cold'' nuclear effects which are not due to QGP formation but that affect $J/\psi$ production nonetheless. Among them, the nuclear modification of the parton densities (nPDF) may play a role in the nuclear dependence of $\jpsi$ production. Another effect is the inelastic rescattering of the $J/\psi$ state in a nuclear matter, the so-called nuclear absorption. It is therefore crucial to have a quantitative understanding of these cold nuclear effects in order to get a quantitative understanding of the $\jpsi$ suppression reported in heavy systems at SPS~\cite{na50,na60} and RHIC~\cite{phenix} and, therefore, a reliable to interpretation of the observed suppression.

Nuclear absorption is expected to be the dominant source of $J/\psi$ suppression not only in peripheral heavy nucleus-nucleus (AA) collisions but also in hadron(photon)--nucleus reactions, which are dominantly sensitive to cold nuclear effects. Experimentally, a large variety of $J/\psi$ production off nuclear targets have been measured at various colliding energies (SPS~\cite{Badier:1983dg,Abreu:1998ee,Alessandro:2003pc}, FNAL~\cite{Katsanevas:1987pt,Kartik:1990it,Alde:1990wa,Leitch:1999ea}, HERA-B~\cite{Husemann:2005yq}, RHIC~\cite{Adler:2005ph}, SLAC~\cite{Anderson:1976hi} and NMC~\cite{Amaudruz:1991sr}). A global analysis of all available measurements of $J/\psi$ production in nuclear target allows for a study of nuclear absorption effects and to quantify the strength of that mechanism, monitored essentially by one physical parameter, the $J/\psi$-nucleon inelastic cross section, $\sig$.
In the following, the extraction of $\sig$ assuming nuclear parton modifications is presented within the framework used in a previous analysis~\cite{arleo_tram}.

In this paper, the most recent measurements performed by the PHENIX experiment in $d$--Au collisions~\cite{phenix_dAu_new} are analyzed. Moreover, additional results using the EPS08 nPDF parametrization~\cite{eps08} are given.

\section{Extracting $\sig$}
\label{sec:method}

This section gives a brief description of the method followed in this analysis: the model used to describe the data selected, the nuclear parton distribution implementation, the data sets and finally the fitting method. A more detailed description of the method can be found in~\cite{arleo_tram}. 

\subsection{Model}
\label{sec:model}
The various $J/\psi$ production channels in the different reactions in the data sample are the following:
\begin{equation}
  \label{eq:reactions}
(p, \bar{p}, \pi^+, \pi^-, \gamma^*) + \ A \rightarrow  J/\psi + \X 
\end{equation}
The $J/\psi$ production cross section $\sig^{\produc}$ in hadronic collisions is determined within the Colour Evaporation Model~\cite{Barger:1979jsBarger:1980mg}  (CEM) at leading order (LO). The PDF in the hadron projectiles are taken from the LO parametrization SMRS for the pion~\cite{Sutton:1991ay} and CTEQ6L for the (anti)proton~\cite{Pumplin:2002vw}. Since only cross section {\it ratios} are analyzed in the following, the results from this analysis are almost independent on the specific choice of the proton PDF parametrization.

The survival probability $S_{\rm{abs}}(A,\sig)$ of $\jpsi$ states propagating in a nucleus A --~i.e. the probability for no inelastic interaction~-- is given in the Glauber model by~\cite{Capella:1988ha}
\begin{eqnarray}
  \label{eq:supp}
S_{\rm{abs}}(\A, \sig)& =& \frac{1}{(A-1) \ \sig}\nonumber\\ &\times& \int \dd {\bf b} \left( 1 - e^{- (1-1/A) \ \TA({\bf b}) \ \sig} \right).
\end{eqnarray}
with the thickness function $\TA({\bf b})$ 
\begin{equation}
  \label{eq:tf}
  \TA({\bf b}) = \int_{-\infty}^{+\infty} \dd z \ \rho({\bf b}, z).
\end{equation}

It depends on the atomic mass number $A$ of the nucleus and the $J/\psi$--N inelastic cross-section, $\sig$. The observed $J/\psi$ production as a function of the longitudinal momentum fraction $x$, then is:
\begin{equation}
  \label{eq:xs}
  \frac{\dd \sig }{\dd x} = S_{\rm{abs}}(\A\ , \sig) \times \frac{\dd \sig^{\produc} }{\dd x},
\end{equation}
In this current analysis, the cross section ratios $R^{th}$ of heavy (A) to light (B) nuclei are considered:
\begin{equation}
 \label{eq:Rth}
R^{\rm th} (\sig) = \frac{B}{A} \frac{\dd\sigma (h,\gamma^* \A \rightarrow  J/\psi X)/\dd{x} }{\dd\sigma(h,\gamma^* \B \rightarrow  J/\psi X)/\dd{x}}
\end{equation}
Note that since only ratios of cross sections at the same energy are used, most uncertainties regarding the $\jpsi$ production cross sections cancel.

It is worthwhile to note that formation-time effects are neglected, in the sense that the question of which state actually propagates through the nuclear matter is not addressed. Also, the feed down from higher mass resonances is not taken into account. Consequently, $\sig$ has to be seen as an effective absorption parameter resulting from an average of the $c\bar{c}$ and $J/\psi$, $\chi_c$ and $\psi^\prime$ interaction with nucleons, rather than the genuine $\jpsi$--N inelastic cross section. 

\subsubsection{Nuclear parton distributions}
\label{sec:npdf}
Partons in bound nucleons show noticeably different momentum distributions as compared to those in free protons. This modification is quantified by $R(x,Q^2,A)$ as a function of Bjorken variable, the square of the momentum transfer $Q^2$ and the nucleus size A in the following formula~:
\begin{eqnarray}
R_i(x,Q^2,A)  = f_i^{\A}(x,Q^2)\ /\ A f_i^p(x,Q^2)
\label{eqn:gluon_mod}
\end{eqnarray}
where $f_i$ and $f_i^\A$ describe respectively  the distribution of parton $i$ in a proton and in a nucleus.

\begin{figure}[htbp!]
\resizebox{0.45\textwidth}{!}{%
  \includegraphics{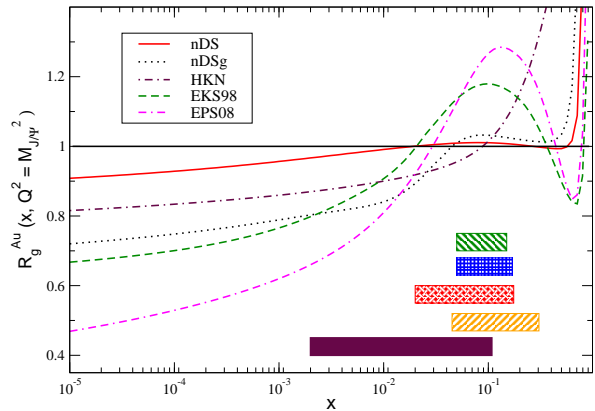}
}
\caption{The ratio of the gluon distribution in a gold nucleus over that in a proton, $R_g^{\rm Au}(x, M_{J/\psi}^2$), plotted as a function of the Bjorken variable using the nDS, nDSg~\cite{deFlorian:2003qf}, EKS98~\cite{Eskola:1998iyEskola:1998df}, HKN~\cite{Hirai:2001npHirai:2004wq} and EPS08~\cite{eps08} parametrizations. The bands indicate the typical $x$-range probed by $\jpsi$ production in the NMC, SPS, FNAL, HERA-B, and RHIC experiments (top to bottom).}
\label{fig:gluon_shadowing}
\end{figure}

Since $\jpsi$ is predominantly produced via gluon fusion in $p$--A collisions\footnote{In $\pi^\pm$--A and $\bar{p}$--A collisions, the scattering of a valence antiquark from the projectile with a valence quark from the target is favoured.} at high energy ($\sqrt{s_{p\A}}>$20~GeV) its production is affected by the modification of the gluon distribution in nuclei. Several DGLAP analyses~\cite{deFlorian:2003qf,Hirai:2001npHirai:2004wq,Eskola:1998iyEskola:1998df} aim at the extraction of the ratios $R_i(x, Q^2, \A)$ from DIS and Drell-Yan data. However, given the indirect constraints in the gluon sector (through scaling violations), the ratio $R_g$ is poorly determined. Figure~\ref{fig:gluon_shadowing} shows the gluon distribution in a Au nucleus with various parametrizations available as a function of $x$.

The shaded band area shows the kinematic range of the $J/\psi$ production (at LO) for various experiments, NMC (green), SPS (blue), FNAL (red), HERA-B (orange) and RHIC (purple), from top to bottom. One can observe that $J/\psi$ production is affected by mainly two effects, the anti-shadowing ($R^{\rm Au}_g >1$ at $2$--$5\times10^{-2}\lesssim x\lesssim 0.3$) for SPS, FNAL and HERA-B and the shadowing effect ($R^{\rm Au}_g<1$ at $x\lesssim 10^{-2}$) at RHIC. A strong anti-shadowing effect increases $J/\psi$ production in nuclei with respect to the (binary scaled) production in $p$--$p$ collisions, leading to a cross-section ratio larger than 1. This enhanced production will in turn be compensated by an increase of the fitted nuclear absorption cross section (SPS, FNAL and HERA-B). Conversely, a strong shadowing effect tends to deplete the nuclear absorption cross section (RHIC energy).

In this work, the EPS08 parametrization (magenta, dotted-dashed-dashed line) is added in the analysis. This nPDF set exhibits  a strong anti-shadowing effect in comparison with the previous EKS98 distributions, and the anti-shadowing region is also slightly shifted to higher $x$ values ($2\times10^{-2}<x<0.3$). 
In addition, the shadowing effect is much stronger than in EKS98, due to the inclusion in the analysis of these authors of the recent RHIC data.

\subsection{Data sets}
\label{sec:data}

Since factorization between $\jpsi$ production and the subsequent inelastic interaction is assumed in the present analysis, both hadroproduction (using pion, proton, antiproton and deuterium beams) and leptoproduction data are analyzed.

The detailed data selection list can be found in~\cite{arleo_tram}. Concerning hadroproduction measurements, the projectiles used were mainly protons (NA3, NA38, NA50, E866, HERA-B, PHENIX), but also pions (E537, NA3, E672), antiprotons (E537), and deuterium nuclei (PHENIX). The range of colliding energy is $\sqrt{s_{NN}}=15$--$200$~GeV.

As mentioned previously, PHENIX data have been reanalyzed~\cite{phenix_dAu_new}, the measurements in $d$--Au are now normalized with respect to higher-statistics $p$--$p$ collisions measurements performed at positive/negative rapidity instead of an average of measurements at positive/negative rapidity. The $R^{\rm exp}$ ratios are now smaller than in the previous analysis~\cite{phenix} for each rapidity region.
Concerning the uncertainties, the precise $p$--$p$ measurements also lead to slightly smaller statistical uncertainties. However, since the data are not taken during the same year and with the same configuration, the systematic uncertainties which used to cancel in the ratio are now larger.

As for leptoproduction experiments, the NMC data~\cite{Amaudruz:1991sr} are selected. The virtual-photon energy $\nu$ ranges from 40 to 240~GeV in the laboratory frame, corresponding to $\gamma^*$--N centre-of-mass energies $\sqrts=8$--21~GeV and the Bjorken-$x$ range probed in the gluon distributions of the nuclear targets is $x=0.05$--$0.15$ in the NMC kinematics.

After the data selection, the $R^{\rm exp}$ ratios of heavy (A) to light (B) nuclei are determined and compared to the $R^{\rm th}$ ratios. In order to avoid too large systematic errors in the experimental ratio, both reactions on targets $\A_{_i}$ and B are required to be taken from the same experiment and at the same centre-of-mass energy. For each experiment, the uncertainties on the $R^{\rm exp}$ ratios are then separated as follows:
\begin{equation}
 \label{eq:Rexp}
R^{\rm exp}_i \pm \sigma_i \pm \beta\ R^{\rm exp}_i
\end{equation}
where $\sigma_i$ represents the uncorrelated errors (statistical and the uncorrelated systematic errors, added in quadrature), and $\beta$ corresponds to the normalization correlated error, often coming from the fact that cross sections in different nuclei are {\it all} normalized to the {\it same} light target (hence with an error common to all $R_i$).

\subsection{Fitting method}
\label{sec:fit}

 The $J/\psi$--N inelastic cross section is extracted, for each experimental sample $\ell$ with $n_{_\ell}$ data points, from the minimization of the $\chi^2_{_\ell}$ function~\cite{Stump:2001gu}:
\begin{equation}
  \label{eq:chi2}
  \chi^2_{_\ell}(\sig) = \sum_{i = 1}^{n_{_\ell}} \ \left[\frac{R_i^{\rm exp} - R_i^{\rm th}(\sig)}{\sigma_{i}}\right]^2 - V^2/M,
\end{equation}
computed from Eq.~(\ref{eq:Rth}), depends explicitly on the free, but positive, parameter, $\sig$. The {\it correlated} normalization error $\beta$, on the data point $i$ enters the $V$ and $M$ in Eq.~(\ref{eq:chi2}) through:
\begin{eqnarray*}
V \ &=&\ \sum_{i=1}^{n} \ \frac{\beta\ R_i^{\rm exp}\ [R_i^{\rm exp} - R_i^{\rm th}(\sig)]}{\sigma^2_i},\\
M \ &=&\ 1 \ +    \beta^2\ \sum_{i=1}^n\frac{(R_i^{\rm exp})^2}{\sigma_i^2}.
\end{eqnarray*}
The $1\sigma$ error $\delta\sig$ on the fitted parameter $\sig$ is defined so as to increase $\chi^2$ by one unit from its minimum: 
\begin{equation}
  \label{eq:dchi2}
  \Delta \chi^2 \equiv \chi^2(\sig \pm \delta \sig) - \chi_{_{\rm min}}^2 = 1.
\end{equation}

\section{Determination of $\sig$ from each experiment}
\label{sec:results_01}
The analysis using new PHENIX results are compared to the previous analysis in Table~\ref{tab:phenix}. Since the $R^{\rm exp}$ ratios in the PHENIX new analysis~\cite{phenix_dAu_new} are smaller than in~\cite{phenix}, the nuclear absorption cross section obtained in this work is now higher than previously, by roughly 2~mb.  The cross sections now vary from 2.5 $\pm2.2$~mb (with nDSg) to $5.4\pm2.5$~mb (proton PDF) with various nPDF parametrizations.

\begin{table*}[htb] 
  \caption{The $\jpsi$--N cross section extracted from the new re-analyzed PHENIX data versus previous analysis using the proton and various nuclear parton density parametrizations. The $\chi^2/\ndf$ and the $\chi^2$ probability are also shown.}
 \centering
{\small
\begin{tabular}[c]{p{1.7cm}|ccc|ccc|c}
    \hline
 & \multicolumn{3}{c}{Previous results using~\cite{Adler:2005ph}} &        \multicolumn{3}{c}{New analysis using~\cite{phenix_dAu_new}}& Absolute change\\
    & $\sig$ (mb)& $\chi^2/\ndf$ & Probability& $\sig$ (mb)& $\chi^2/\ndf$ &Probability&\\
    \hline
 & & & & & & & \\
 proton      &  3.5 $\pm$  3.0 &  1.7 & 0.79 &  5.4 $\pm$  2.5 &  0.84& 0.93& $+1.9$~mb \\
 nDS       &  3.1 $\pm$  2.6 &  1.4 & 0.84 &  5.1 $\pm$  2.5 &  0.69&0.95& $+2.0$~mb\\
 nDSg      &  0.6 $\pm$  1.9 &  0.8 & 0.93  & 2.5 $\pm$  2.2 &  0.27& 0.99& $+1.9$~mb\\
 HKN      &  1.5 $\pm$  2.3 &  1.3 &  0.86 &  3.2 $\pm$  2.3 &  0.56 & 0.97   & $+1.7$~mb\\
 EKS98      &  1.3 $\pm$  2.0 &  0.6 &  0.93 & 3.1 $\pm$  2.2 &  0.12 & 1.00 & $+1.8$~mb\\
 EPS08      &  1.3 $\pm$ 2.5  & 1.5 & 0.83 &  2.2 $\pm$  2.2 &  0.37&0.98 & $+0.9$~mb\\
 & & & & & & & \\
    \hline
\end{tabular}}
\label{tab:phenix}
\end{table*}

The extracted nuclear absorption cross sections using the EPS08 nPDF parametrization are shown in Table~\ref{tab:results_shadowing_all} for all individual experiments. These results are compared with the results obtained previously in~\cite{arleo_tram} using the EKS98 parametrization.
\begin{table*}[htb] 
\caption{The $\jpsi$--N inelastic cross section, $\chi^2/\ndf$ extracted from each data sample using EKS98 and EPS08 parametrizations for the nuclear PDFs.}
 \centering
{\small
\begin{tabular}[c]{p{1.7cm}|cc|cc|c}
    \hline
   Exp. & $\sigEKS$ (mb)& $\chi^2_{_{\rm EKS}}/\ndf$ & $\sig^{\rm EPS}$ (mb)& $\chi^2_{_{\rm EPS}}/\ndf$ & Relative change\\
    \hline
 & & & & & \\
 E537      &  8.2 $\pm$  1.1 &  1.9 &    9.0 $\pm$  1.2 &  1.86& $+10\%$\\
 NA3       &  4.6 $\pm$  0.2 &  1.2 &    5.2 $\pm$  0.2 &  1.32& $+13\%$ \\
 NA38      &  7.9 $\pm$  0.8 &  3.2 &    9.0 $\pm$  0.8 &  3.07& $+14\%$\\
 NA50      &  6.8 $\pm$  0.5 &  0.3 &    7.8 $\pm$  0.5 &  0.31   & $+15\%$\\
 E672      &  11.6 $\pm$  6.3 &  0.6 &  10.0 $\pm$  5.8 &  0.61 & $-14\%$\\
 E866      &  5.3 $\pm$  1.7 &  6.5  &  8.0 $\pm$  3.7 &  20.4 & $+51\%$\\
 HERA-B    &  4.2 $\pm$  1.5 &  0.9 &    5.1 $\pm$  1.5 &  0.8   & $+21\%$\\
 PHENIX    &  3.1 $\pm$  2.2 &  0.12 &    2.2 $\pm$  2.2 &  0.37 & $-29\%$\\
 NMC       &  $\le$ 1.6 &  0.5 &  $\le$  2.00 &  0.35 & $+25\%$ \\
 & & & & & \\
    \hline
\end{tabular}}

  \label{tab:results_shadowing_all}
\end{table*}
Because of the more pronounced shadowing in the EPS08 parametrization than in EKS98, the extracted nuclear absorption cross section at RHIC energy decreases by $\sim30\%$. The anti-shadowing effect is also more pronounced in the EPS08 parametrization, leading to an increase by roughly $10$--$20\%$ of the nuclear absorption cross section at the energies of the SPS, FNAL~\footnote{The result obtained using the E672 measurements does not exhibit this increase, the x region probed ($x \sim$0.03) lies in the anti-shadowing region in EKS parametrization while within the EPS parametrization, this x region is at the boundary of the shadowing region with a weak effect of the nuclear distribution.} and HERA-B experiments. Finally, note the significant increase $(+50\%)$ from EKS98 to EPS08 using the E866/Nusea data samples. However the large $\chi^2/\ndf\simeq 20$ obtained is large, as stressed in~\cite{arleo_eps}.

The nuclear absorption cross section for PHENIX depends on the strength of shadowing in each nPDF set: a strong shadowing parametrization leads to a decrease of the nuclear absorption cross section. On the contrary, when using a proton PDF or a nPDF with no (or weak) shadowing effect, the nuclear absorption cross section is then higher to compensate for the (weak) suppression due to gluon shadowing.

\begin{figure}[htbp!]
\resizebox{0.45\textwidth}{!}{%
  \includegraphics{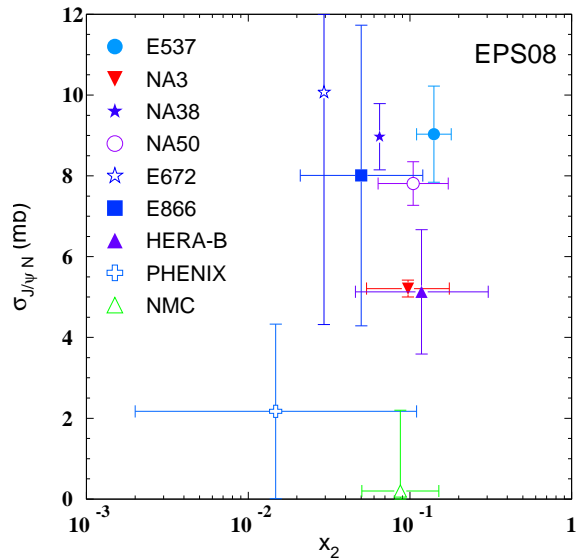}
}
\caption{The $\jpsi$--N cross section extracted from each data set, using EPS08 ($\sigEPS$) nuclear parton densities as a function of $x_2$.}
\label{fig:eks_x2}
\end{figure}

Since the energy in the $\jpsi$-nucleon (or $c\bar{c}$-nucleon) system, given by $\sqrtspsin \simeq \mpsi/ \sqrt{x_{_2}}$, is directly related to the momentum fraction $x_2$, one could expect the extracted $\sig$ cross section to be a scaling function of $x_2$. As discussed in~\cite{arleo_tram}, there is no real $x_2$-dependence observed within this framework using a proton distribution, or using nDS, nDSg, EKS98 and HKN nuclear parton distribution. For completeness, Figure~\ref{fig:eks_x2} shows the nuclear absorption cross section $\sig$ as a function of $x_2$ using the EPS08 nPDF. In the region of $x_2 \sim$ 0.1, one can observe that the spread of extracted $\sig$ reported using the other nPDF sets persists. Interestingly, it also appears that using EPS08 leads to some decrease of $\sig$ from fixed-target to RHIC energies, indicating possible formation-time effects at small $x_2$. Also, in the previous analysis~\cite{arleo_tram}, a similar trend has been observed when using EKS08 parametrization. However, the error bars are too large to make any firm conclusion for EKS08/EPS08 and other nPDF sets.
Note that higher-twist production mechanisms may very well have a different kinematic dependence; this is for intance for the case for the intrinsic charm model which naturally exhibits a Feynman-x scaling (see e.g.~\cite{brodsky}). However, we expect its contribution to be small at low $|x_F|$ which we consider here.

\section{Global fit and discussions}

In the following, a global fit is performed assuming that the $\sig$ dependence on energy is weak. The detailed method for the global fit is described in~\cite{arleo_tram}. The $1\sigma$ error is rescaled,
\begin{equation}
\delta \sigb \ = \ {\rm S} \ \times \ \delta \sig,
\end{equation}
where the factor S is defined by:
\begin{equation}\label{eq:Sfactor}
{\rm S} \ \equiv \sqrt{\frac{\chi^2}{n-1}} \quad {\mathrm{if}} \quad \chi^2/\ndf > 1, 
\end{equation}
with $n$ data points and S~$\equiv 1$ otherwise.
The $\jpsi$--N cross section is systematically determined from the individual data samples. \\
The extracted $\sig$ is then determined from the minimization of the weighted $\chi^2$ function:
\begin{equation}
  \chi^2(\sig) = \sum_{\ell=1}^{\cal N} \ \ {\rm S}_{_\ell}^{-1} \ \ \chi_{_\ell}^2(\sig).
\end{equation}
with the individual $\chi_{\ell}$ for each experimental data sample.
This global fit analysis will thus favour data sets with a small individual $\chi^2/\ndf$. The results obtained from this global fit using a proton PDF and various nPDF parametrizations (nDS, nDSg, HKN, EKS98 and EPS08) are summarized in Table~\ref{tab:shad}. These results include the recent PHENIX results~\cite{phenix_dAu_new} already mentioned. The $\chi^2/\ndf$ from these fits varies from 1.4 to 1.7.

The spread of $\sigsh$ quoted in Table~\ref{tab:shad} directly reflects the present lack of knowledge of the (gluon) nuclear densities. Taking the nDS parametrization as the default set, the cross section extracted in this analysis is:
\begin{equation}
 \sig=3.5\pm 0.2\ ({\rm stat.})\pm 2.6\ ({\rm syst.})~\rm{mb}, 
\end{equation}
where the systematic error quoted here only comes from the uncertainties of the nPDFs. Clearly, a better determination of $\sigsh$ could only be achieved when these are more tightly constrained by experimental data. 

\begin{table*}[htb]
  \caption{The $\jpsi$--N cross section extracted from the data using the proton and various nuclear parton density parametrizations.}
  \centering
  \begin{tabular}[c]{ccccccc}
\hline
   & Proton& nDS & nDSg & EKS98 & HKM &EPS08 \\
\hline
$\sigsh$ (mb) & 3.4 $\pm$ 0.2 & 3.5 $\pm$ 0.2 & 4.0 $\pm$ 0.2 & 5.2 $\pm$ 0.2 & 3.6 $\pm$ 0.2 & 6.0 $\pm$ 0.2\\
$\chi^2/\ndf$ & 1.4 & 1.4 & 1.5 &  1.5 &  1.4 & 1.7\\
\hline
  \end{tabular}

  \label{tab:shad}
\end{table*}

\begin{figure}[htbp!]
\resizebox{0.45\textwidth}{!}{%
  \includegraphics{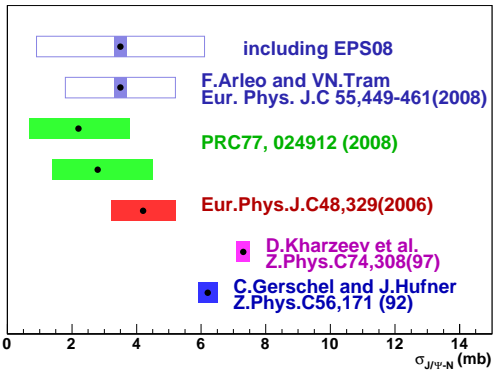}
}
\caption{The $\jpsi$--N cross section extracted in the global fit in this work compared to previous analyses.}
\label{fig:fit_comparison}
\end{figure}

Figure~\ref{fig:fit_comparison} shows the fitted $\sig$ from this work compared to another global analysis (Gerschel and H\"ufner (GH) in~\cite{Gerschel:1993uh}, Kharzeev et al. (KLNS) in Ref.~\cite{Kharzeev:1996yx}) as well as with the extracted $\sig$ by NA50~\cite{Alessandro:2003pc} and PHENIX~\cite{phenix_dAu_new} from their data. Both results by Gerschel and H\"ufner (GH) in~\cite{Gerschel:1993uh} and Kharzeev et al. (KLNS) in Ref.~\cite{Kharzeev:1996yx} are significantly higher than the $\sig$ cross section presented in this work. These differences are believed to be mainly due to the different data sets used in the global analysis. A more detailed discussion can be found in~\cite{arleo_tram}.

Compared to NA50~\cite{Alessandro:2003pc} analysis, the $\sig$ extracted on their measurement is compatible with the individual $\sig$ extracted from this work using a proton parton distribution, namely $\sig^{\rm NA50}=4.2\pm0.5$~mb {\it vs.} $4.7\pm0.5$~mb in this work.
In the new analysis of the PHENIX data~\cite{phenix_dAu_new}, the collaboration also published the value of $\sig$ using the nDSg and EKS98 parametrizations. The results are compatible with results presented in this work within the error bars. When using the EKS98 parametrization, PHENIX results are $2.8^{+1.7}_{-1.4}$ {\it vs.} $3.1\pm2.2$~mb in this work and when using nDSg, PHENIX results are $2.2^{+1.6}_{-1.5}$~mb {\it vs.} $2.5\pm2.2$~mb.

\section{Summary}
\label{sec:summary}

In this work, a re-analysis of the nuclear absorption cross section using the new PHENIX results within the framework described in~\cite{arleo_tram} is presented. The largest suppression reported in the new PHENIX analysis leads to an increase of $\sig$ from $3.5\pm0.3$~mb to $5.4\pm2.5$~mb using the PDF of the proton. The $\sig$ obtained in this work is also compatible within the uncertainties with the value determined from the PHENIX analysis of their measurements. It is worthwhile to note that RHIC has provided high-statistics $d$--Au collisions during the 2008 year data taking, and the analysis of this new set of data should allow for a more precise measurements of the $R^{\rm exp}$ ratios.

In addition, an analysis of the $\sig$ nuclear absorption cross section is performed using the EPS08 nPDF set; it is presented for each individual experiment. The strong shadowing and anti-shadowing effects described by this parametrization induce in turn a possible $x_2$ dependence of the $\sig$ cross section, leading to a smaller nuclear absorption at low $x_2$ (RHIC energy) and increasing it at large $x_2$ (SPS, FNAL, HERA-B). However, the discrepancy of the extracted $\sig$ observed at large $x$ ($x\sim$0.1) persists. Finally, a global fit including the new PHENIX results and the EPS08 parametrization is presented in this work.

\end{document}